# I♥LA: Compilable Markdown for Linear Algebra

YONG LI, George Mason University, USA
SHOAIB KAMIL, Adobe Research, USA
ALEC JACOBSON, University of Toronto and Adobe Research, Canada
YOTAM GINGOLD, George Mason University, USA

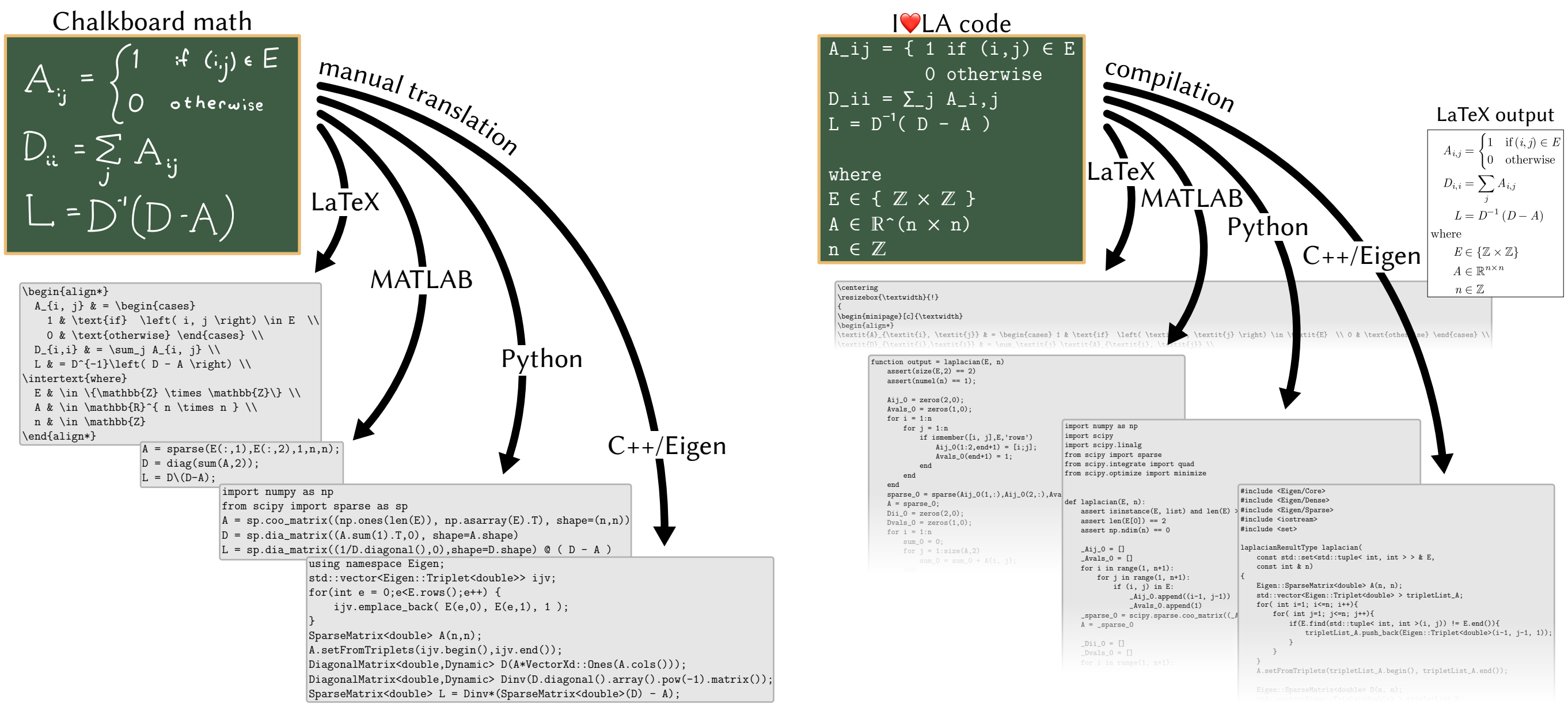

Fig. 1. Left: Mathematical notation has evolved over centuries to efficiently communicate technical concepts such as the sparse graph Laplacian construction in the top left. Meanwhile, programming languages communicate with a machine typically with a reduced character set and syntax causing handwritten translation of mathematics to visually stray far from the "chalkboard math" and from each other. Right: I♥LA is a novel domain specific language for linear algebra. The I♥LA code written with rich unicode symbols visually resembles chalkboard math, while still being a semantically well-defined programming language compilable to various target languages: LaTeX, MATLAB, Python, C++.

Communicating linear algebra in written form is challenging: mathematicians must choose between writing in languages that produce well-formatted but semantically-underdefined representations such as LaTeX; or languages with well-defined semantics but notation unlike conventional math, such as C++/Eigen. In both cases, the underlying linear algebra is obfuscated by the requirements of esoteric language syntax (as in LaTeX) or awkward APIs due to language semantics (as in C++). The gap between representations results in communication challenges, including underspecified and irreproducible research results, difficulty teaching math concepts underlying complex numerical code, as well as repeated, redundant, and error-prone translations from communicated linear algebra to executable code. We introduce I♥LA, a language with syntax designed to closely mimic conventionally-written linear algebra, while still ensuring an unambiguous, compilable interpretation. Inspired by Markdown, a language for writing naturally-structured plain text files that translate into valid HTML, I♥LA allows users to write linear algebra in text form and compile the same source into LaTeX, C++/Eigen, Python/NumPy/SciPy, and MATLAB, with easy extension to further math programming environments. We outline the principles of our language design and highlight design decisions that balance between readability and precise semantics, and demonstrate through case studies the ability for I♥LA to bridge the semantic gap between conventionally-written linear algebra and unambiguous interpretation in math programming environments.

Authors' addresses: Yong Li, George Mason University, USA, yli69@gmu.edu; Shoaib Kamil, Adobe Research, USA, kamil@adobe.com; Alec Jacobson, University of Toronto and Adobe Research, Canada, jacobson@cs.toronto.edu; Yotam Gingold, George Mason University, USA, ygingold@gmu.edu.

## 1 INTRODUCTION

Linear algebra has become the *lingua franca* of computer graphics and other fields of scientific computing. Matrices and vectors allow researchers to succinctly communicate mathematical expressions involving arbitrary amounts of data. Mathematical notation using these constructs is a human language that continues to evolve over time to be readable yet precise, while eliding details unnecessary for communication. It allows for communicating expressions and formulae to other researchers, practitioners, educators, and students, often by publishing scientific papers.

Communicating expressions and formulae requires writing them in a language that produces mathematical notation, such as LaTeX, with esoteric syntax that is not readily understandable without compiling to a print-ready document. On the other hand, *implementing* mathematical expressions entails translating them into a specific programming environment: a language and linear algebra package, with its own specific syntax dissimilar to languages designed for expressing syntax. Thus, succinct expressions must be rewritten multiple times in languages less legible than linear algebra (Fig. 1, left). This has the potential to introduce bugs. Worse, implementers, in both academia and industry, favor different programming environments—and these environments change over time. To participate in research and development, everyone must become a highly proficient implementer, which excludes many from contributing. This leads to an explosion of redundant effort, differing implementations, bugs, and a less diverse community.

To address these issues, we introduce I❤️LA, a language whose syntax is designed to be as close as possible to conventionally written math yet can be written in a plain text editor (Fig. 1, right), all while ensuring an unambiguous, compilable interpretation. To achieve this, we make language design decisions that accord with mathematical convention rather than conventional programming languages. For example, we prioritize single-letter identifiers (possibly with Unicode marks), identifier juxtaposition is multiplication, and we promote Unicode characters (which are as easy to input as TeX commands and no more difficult than typing a function name in a traditional programming language) as operators and variable names.

Our current compiler can output LaTeX, Python/NumPy/SciPy, MATLAB, and C++/Eigen, creating a publishable artifact and reference implementations that guarantee replication, all from the same source file. I❤️LA includes syntax for creating matrices (dense, sparse, and block), for algebraic and linear algebra expressions involving matrices and vectors, for sets and sequences, for summation and integration, and for constrained minimization. I❤️LA's visual similarity to conventional math comes from design decisions that balance three competing objectives: the language must look as close as possible to conventional math notation syntactically; it must be writable in a plain text editor (with Unicode support); and the language must retain unambiguous semantics that can be compiled to common programming languages and linear algebra packages.

In designing I❤️LA, we set out to design a language whose notation matched the notation commonly used in computer graphics papers when communicating linear algebra concepts. While I❤️LA is likely applicable outside of computer graphics, we designed it to solve problems in the graphics community as a first step. I❤️LA provides the ability to write individual formulas, such as those commonly found in computer graphics publications, rather than full applications that use these formulas. Towards this end, we systematically analyzed all numbered equations in *ACM SIGGRAPH* 2019 papers. We pursued syntax-driven development. We carefully considered consensus-syntax for the math we support. We describe the core design decisions and simplifications that make it tractable to create I❤️LA without sacrificing readability, while still supporting conventional notation and clear semantics. We demonstrate the use of I❤️LA through case studies of graphics algorithms that illustrate how the language enables practitioners to write in plain text and use the same source for generating output suitable for publications/communication as well as reference implementations in both statically and dynamically typed languages. We performed a user study to understand how experienced practitioners learn and perceive I❤️LA.

## 2 RELATED WORK

For researchers and practitioners alike to make use of each others' techniques, they must translate the linear algebra expressions into a specific programming environment. Examples today include: Python with NumPy/SciPy or TensorFlow or PyTorch; C/C++ with Eigen or PETSc or Armadillo or BLAS; MATLAB; Fortran; R; and Julia. Some environments are less than 10 years old (Julia, Armadillo, TensorFlow, PyTorch). Some previously popular environments have fallen out of fashion or were updated in a backwards-incompatible manner (TensorFlow 2 versus 1, PyTorch versus Torch, Theano, Python 3 versus 2). Recent domain-specific languages like TACO [Kjolstad et al. 2017], BLAC [Spampinato and Püschel 2014], YALMIP [Löfberg 2004], and GENO [Laue et al. 2019] produce highly efficient computation but also define their own input languages. None of these languages take input as legible, conventionally-written math.

Several programming languages allow users to write expressions that resemble or borrow notation from conventional mathematics in some ways. Notably, Fortress [Allen et al. 2005] also defined juxtaposition as multiplication. Julia [Bezanson et al. 2017] allows mathematical glyphs to be used as infix function operators. We provide a comparison of I❤LA, MATLAB, Mathematica, and Julia in Table 1. Adopting one of these languages requires embracing its ecosystem, whereas I❤LA aims to be ecosystem agnostic. Our technical contribution is a formalization and parser [Ganesalingam 2013] for conventionally-written mathematics (numerical linear algebra). We address how equations are typed by paper writers and others communicating using mathematical notation. The Fortress language has been abandoned, so code written in it can no longer be run. This argues for our meta-compilation approach, which can re-target I❤LA to new output languages as they arise. Languages like Lean [de Moura et al. 2015], Agda [Norell 2007], and Coq [Team 2021] target a very different portion of mathematics (proofs and formal verification). APL [Iverson 2007] introduced new, powerful notation unlike conventional mathematics.

Table 1. A comparison of several languages designed to resemble conventional mathematical notation. This formula was used in Python Enhancement Protocol 465 [Smith 2014] to motivate the introduction of a dedicated matrix multiplication operator. AsciiMath and LaTeX are not executable.

| | |
|---|---|
| I❤️LA | `S = (Hβ-r)ᵀ(HVHᵀ)⁻¹(Hβ-r)` |
| MATLAB | `S = (H*beta-r)'*((H*V*H')\(H*beta-r))` |
| Mathematica | `S = (H.β-r)ᵀ.LinearSolve[(H.V.Hᵀ),(H.β-r)]` |
| Julia | `S = (H*β-r)'*((H*V*H')\(H*β-r))` |
| Python/NumPy | `S = (H@β-r).T@solve((H@V@H.T),(H@β-r))` |
| AsciiMath | `S = (Hbeta-r)^TT(HVH^TT)^(-1)(Hbeta-r)` |
| LaTeX | `S = (H\beta -r)^\top (HVH^\top )^{-1}(H\beta -r)` |

Lightweight markup syntaxes have become enormously popular for prose and mathematics. I❤LA is the first that can be compiled and executed. Markdown [Gruber and Swartz 2004] is an almost ubiquitous markup language for verbal language text. It is designed to be easy to read and write, matching existing plain text conventions where possible, yet outputs to markup languages with a more explicit syntax like HTML. Its widespread adoption is a testament to its ease of use and the benefits that the structured output provide (enhanced formatting and separation of style from content). We are inspired by the ease and popularity of Markdown, and wish to create a similar language for mathematics, with important benefits to the structured output (working algorithms for scientific replicability).

There are several markup languages for writing mathematical notation in plain text, such as LaTeX [Goossens et al. 1994], AsciiMath [Jipsen 2005], and MathML [W3C 2016]. These languages have seen widespread adoption. They are designed to allow users to input all mathematical notation. However, they do not consider the interpretation of the equations themselves. An expression written using mathematical notation expresses a logical statement. In general, such statements do not specify operations to be performed. Perhaps surprisingly, mathematical notation may still be ambiguous or require careful work or additional context to specify in sufficient detail to correspond to a compilable algorithm [Ganesalingam 2013]. Mathematical expressions by themselves do not contain information about the data types of the symbols (e.g., which variables are scalars, vectors, or matrices). When interpreting plain text input, "sin" can refer to the "sine" function or to the product of three scalars "s", "i", and "n". In I❤LA, we aim to define a syntax that is as lightweight and similar to conventionally written math as possible—containing virtually no markup meta-symbols—for a subset of mathematical notation that can be unambiguously interpreted and compiled. Our project to formalize mathematical notation shares motivation with Leslie Lamport's quest to formalize modern proof writing [2012].

Domain-specific languages that aim to make computer graphics practitioners more productive have a long history, making it easier to express shading calculations [Hanrahan and Lawson 1990; He et al. 2018; Perlin 1985], simulations [Bernstein et al. 2016; Kjolstad et al. 2016], non-linear optimization problems [Devito et al. 2017], high-performance routines [Hu et al. 2019; Ragan-Kelley et al. 2012; Yang et al. 2016], integration [Bangaru et al. 2021], and diagrams [Ye et al. 2020]. Several type systems and languages have been proposed to reduce coordinate system and unit errors in graphics [Geisler et al. 2020; Ou and Pellacini 2010; Preussner 2018]. None of these languages supports conventional mathematical notation as we do.

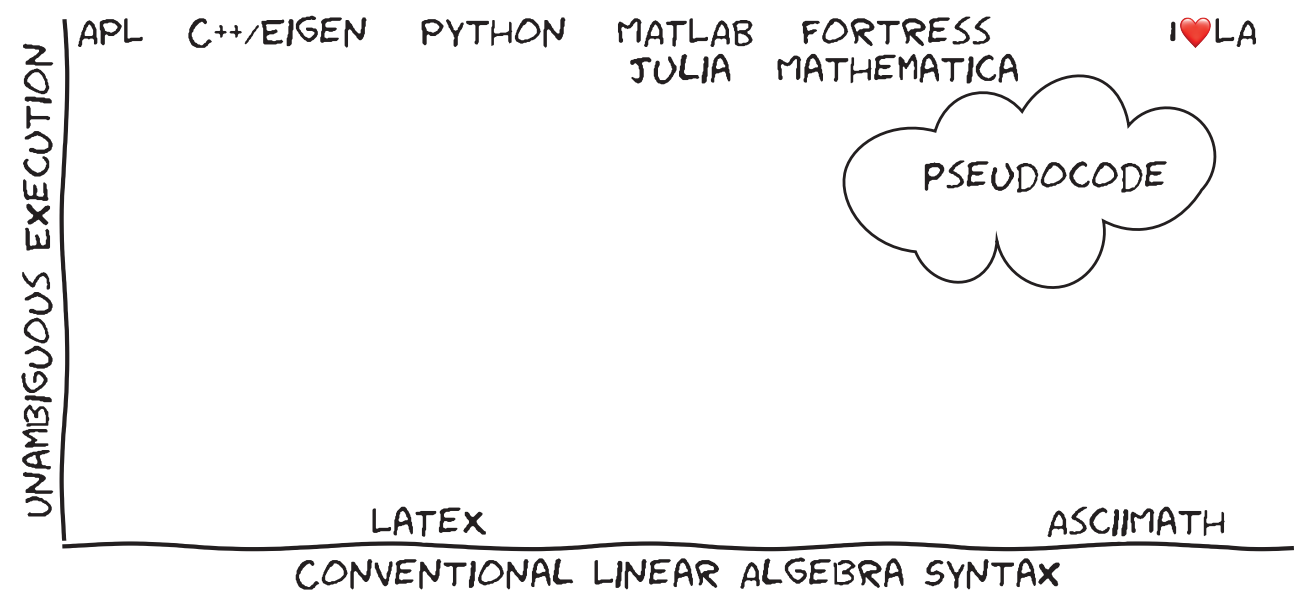

Fig. 2. I❤LA combines conventional syntax with unambiguous execution. Existing environments for inputting linear algebra notation do not consider the interpretability of the expressions. Existing programming languages can be unambiguously compiled, but use a syntax quite unlike conventional mathematical notation. Pseudocode, while readable, cannot be compiled or validated.

## 3 DESIGN OVERVIEW

I❤LA is designed to balance expressivity with guaranteed-correct baseline implementations of formulas fast enough for testing and evaluation. We choose to target individual functions, as usually described inline in research papers and textbooks, eschewing functionality required for creating stand-alone programs. Thus, I❤LA excludes constructs such as file input and output or complex control flow.

The design of I❤LA's syntax and semantics required us to navigate a path between conventional mathematical notation—what one would write on a chalkboard or find in a paper—and the need to unambiguously interpret and compile the notation into machine executable statements (in another programming language). For example, mathematical notation is often used to encode logical steps in service of a derivation or proof. For these expressions, there is nothing obvious to compute. Mathematical notation is also visually complex, involving a wide variety of glyphs in complex arrangements, such as subscripts, superscripts, and the 2D arrangement of matrix elements. Furthermore, mathematical notation is context-dependent, with $f'$ sometimes indicating the derivative of a function $f$ and sometimes purely as a decoration ("f prime"), or inconsistent, such as $\sin^2(x) = \sin(x)\sin(x)$ whereas sometimes $f^2(x) = f(f(x))$. This is in contrast to programming languages, which have historically assumed linear ASCII input, scalar types, and a simple set of operators. Yet the prevalence of linear algebra eventually forced Python, a mainstream programming language, to adopt a dedicated matrix multiplication symbol for a cleaner presentation of expressions [Smith 2014].

We have designed the syntax of I❤LA to follow conventional linear algebra notation while still being expressible in a plain text editor and compilable (Figs. 1 & 2). I❤LA's readability comes from design decisions we make to keep the syntax as close to written math as we can while remaining unambiguously parseable.

To motivate our decisions and inform our choice of features beyond core operations, we tabulated all numbered equations and some unnumbered equations in all technical papers published at *ACM SIGGRAPH* 2019. Out of the 1987 total equations:

- 98% use only single-letter variables (often with decorations),
- 50% use externally defined functions,
- 23% use summations ($\sum$),
- 14% use various norms (e.g., $\|Ax - b\|^2$),
- 10% use minimization (min),
- 10% use common trigonometric functions, and
- 8% use integrals ($\int$).

We used these observations to guide our design decisions, and further refined syntax and expanded the list of operators based on our experiences using I❤LA to implement common operations found in computer graphics, physical simulation, image processing, and geometry processing, as well as general scientific computing.

In Section 4, we illustrate some notable design decisions through examples, including:

(1) Juxtaposition is multiplication. This is the mathematical convention.
(2) Single-letter identifiers are encouraged. This eliminates ambiguity with multiplication and eliminates the need for commas, e.g., when accessing matrix elements with $ij$ subscripts.
(3) Compatible matrix and vector dimensions are statically checked, even when they are parameters rather than fixed.
(4) Matrices can be formatted with 2D elements separated by spaces and newlines.
(5) Variables cannot be re-defined, sidestepping the potential ambiguity between mathematical equality and variable assignment.
(6) Mathematical symbols (operators and variables and marks) are written using Unicode, which makes I❤LA look like conventional mathematics. Unicode is ubiquitous. Modern text editors and operating systems support convenient methods to enter Unicode (e.g., Vim digraphs, Emacs TeX and Agda [Norell 2007] input modes, macOS system-wide text substitutions). Inputting Unicode is no more difficult than inputting LaTeX or typing descriptive variable or method names. Our GUI performs ASCII-to-Unicode substitution[1].

I❤LA programs are pure text files and can be written in a user's preferred text editor. We have additionally developed a simple editor with functionality to replace ASCII with Unicode symbols, syntax highlighting, and the ability to view output in multiple languages simultaneously. As initial output languages, we chose LaTeX for inclusion in papers, two dynamically typed languages (Python with NumPy/SciPy and MATLAB), and a statically typed language (C++/Eigen).

## 4 A TOUR OF THE LANGUAGE

We illustrate the design decisions underpinning I❤LA via a series of examples. I❤LA programs are pure text files. Each I❤LA file compiles to a single function in a target programming language. The following I❤LA code compiles to a function that computes the product of three matrices followed by a quadratic form:

```
given
A ∈ ℝ^(3×n)
B ∈ ℝ^(n×m)
C ∈ ℝ^(m×2)
x ∈ ℝ²

D = ABC
c = xᵀDᵀDx
```

An I❤LA program consists of mathematical statements declaring variable types (e.g., `A ∈`, etc.) or defining variable values (e.g., `D =`, `c =`). Variables cannot be redefined; single-static assignment sidesteps the potential ambiguity between mathematical equality and variable assignment. There is no control flow. The compiled function returns a struct containing all defined variables (e.g., `D` and `c`) for C++, and an object with named members in Python. The last statement in an I❤LA program is always returned, even if it is anonymous (in the member named `ret`). Undefined variables (e.g., `A`, `B`, `C`, `x`) become parameters to the compiled function. Type declarations are inferred in general and need only be specified for input parameters (under the heading `given` or `where`).

In I❤LA, juxtaposition is multiplication. Juxtaposition compiles to the appropriate multiplication: matrix-vector, matrix-matrix, matrix-scalar, scalar-scalar, or scalar-vector, depending on the types of the operands. I❤LA does support using `⋅` but does not support using an asterisk (*) for multiplication, since it is not used in conventional mathematics notation [Cormullion 2020]. The parser correctly interprets the order of operations for the expression `A(y)²` depending on whether `A` is a function or a matrix. To do this, the compiler requires two passes, a first to process type definitions and variable names, and a second to generate the output.

Because I❤LA encourages single-letter identifiers—as in conventional math—there is no ambiguity parsing `ABC` or `xᵀDᵀDx` as multiplication. Multi-letter identifiers are allowed, but can be confusing—just as in conventional math. See the language reference supplemental materials for details. I❤LA identifiers can use all Unicode characters and Unicode marks except a single escape character, backtick (`). Backticks can be used to disambiguate otherwise ambiguous identifiers, such as `` `w_smoothness` ``.

The type checker verifies the compatibility of matrix, vector, and sequence dimensions in expressions. For example, `AC` will report an error because `A ∈ ℝ^(3×n)` while `C ∈ ℝ^(m×2)`, *even though n and m are both arbitrary sizes to be discovered at run-time.* For statically-typed languages, `D` will be a statically-sized $3 \times 2$ matrix.

Vectors are columns. Matrix-vector products (e.g., `Dx`) produce a vector. A vector-matrix product (e.g., `xD`) would raise an error. The transpose of a vector is a row matrix (e.g., `xᵀ`); a row matrix times a vector produces a scalar, so that `xᵀDᵀDx` is equivalent to the dot product `(Dx)⋅(Dx)`.

*Closest Point.* The following I❤LA code compiles to a function that computes the closest point $q$ to a set of 3D lines:

[1] In many cases, I❤LA encourages but does not require Unicode. For example, `∑` can also be written `sum` and superscripts or subscripts can be written `^2` or `_2` instead of `²` or `₂`.

```
given
p_i ∈ ℝ³: points on lines
d_i ∈ ℝ³: unit directions along lines

P_i = ( I₃ - d_i d_iᵀ )
q = ( ∑_i P_i )⁻¹ ( ∑_i P_i p_i )
```

The data types I❤LA supports are can be real `ℝ` or integer `ℤ` scalars; matrices or vectors of scalars; functions taking and returning any of the above; sequences of any of these; or sets of cartesian products of scalars. A variable declared with one subscript is a sequence. In this example, `p_i` and `d_i` are sequences of 3D vectors, and `P_i` is a sequence of $3 \times 3$ matrices. `I₃` is a $3 \times 3$ identity matrix.

Summation in I❤LA takes its bounds from the use of the index in the summand. In this example, it iterates over the elements of `P_i` and `p_i` in the `∑_i` expressions. (I❤LA alternatively supports a conditional expression for the summation index.) The summation operator is conservative rather than greedy. It only sums the first term to its right, not additional terms separated by addition or subtraction. There is some ambiguity in conventionally written math on this matter. For example, is the expression $\sum_i a_i \cdot b_i + c$ identical to $(\sum_i a_i \cdot b_i) + c$ or $\sum_i (a_i \cdot b_i + c)$? This ambiguity grows in complexity if there are additional terms also involving summation, such as $\sum_i a_i + c + \sum_j b_j$ or $\sum_i a_i + c_i + \sum_j b_j$. We considered complex rules, such as inspecting whether the summation index appears in later terms. For guidance, we examined 10 complex summation formulas found in the *ACM SIGGRAPH* 2019 proceedings. Of these, 9 were interpreted correctly with conservative summation. The Wikipedia *Summation* article and a Google image search for "summation" turned up relatively few instances of greedy summation. Based on this evidence, we adopted conservative summation, avoiding the need for I❤LA users to consider complex and possibly surprising behavior.

I❤LA allows comments for type declarations, but not comments in general. I❤LA is designed to support notation-heavy expressions and type definitions in papers, not *entire* papers. It is a future goal to turn I❤LA into a literate programming environment.

*Matrix Examples.* I❤LA supports sparse and dense matrices. Matrices can be defined via linear algebra expressions,

```
A = N⁻¹Mᵀ
```

directly,

```
B = [2a  0
      3 k+1]
```

as block matrices,

```
C = [I  M+yxᵀ
     Mᵀ 0]
```

or element-wise, as in

```
D_ij = M_ij + 7y_i
```

or

```
L_ij = { 1 if (i,j) ∈ E
         0 otherwise
L_ii = -∑_(j for j != i) L_i,j
where
E ∈ { ℤ×ℤ }
L ∈ ℝ^(n×n)
n ∈ ℤ
```

The last example requires a type declaration, since the elements of $L$ are defined conditionally; the dimensions of $L$ cannot otherwise be determined, since there may be additional 0's to the right or below the elements of $E$. In contrast, the identity matrix `I` and zero matrix `0` in the block matrix example `B` have their dimensions unambiguously inferred from other matrices in the row and column. Spaces (not commas) separate elements in matrix declarations. Commas are not needed to separate matrix subscripts (e.g., `L_ij`).

I❤LA automatically creates sparse matrices for scenarios which could possibly benefit. I❤LA's philosophy is to make sensible default choices to keep from burdening the formula writer, even if this leads to non-optimal performance. Matrices are sparse when defined element-wise with a condition, when any matrix in a block matrix definition is sparse, when adding or multiplying two sparse matrices, or when `sparse` is appended to a type declaration. Otherwise, matrices are dense.

*Integration and Minimization.* I❤LA has syntax for simple integration:

```
∫_0^3 ∫_[1, 2] xy ∂x ∂y
```

and simple constrained function minimization:

```
argmin_(x ∈ ℝ³) 1/2 xᵀQx + qᵀx
s.t.
‖x‖ > 1
where
Q ∈ ℝ^(3×3)
q ∈ ℝ³
```

Code generation for these functions is dependent on availability in the target platform. It is out of scope for I❤LA to provide implementations for target environments that do not already include numerical integration and function minimization. I❤LA can be used to target mathematical modeling languages in the future [Dunning et al. 2017; Grant and Boyd 2014; Laue et al. 2019; Löfberg 2004].

*More Examples.* I❤LA also supports common operators (e.g., dot product, cross product, norms, Kronecker product, Hadamard product, inverse and backslash division) and a set of built-in functions and constants (trigonometry and linear algebra). See the examples from the wild (Section 6) and the supplemental language reference for details.

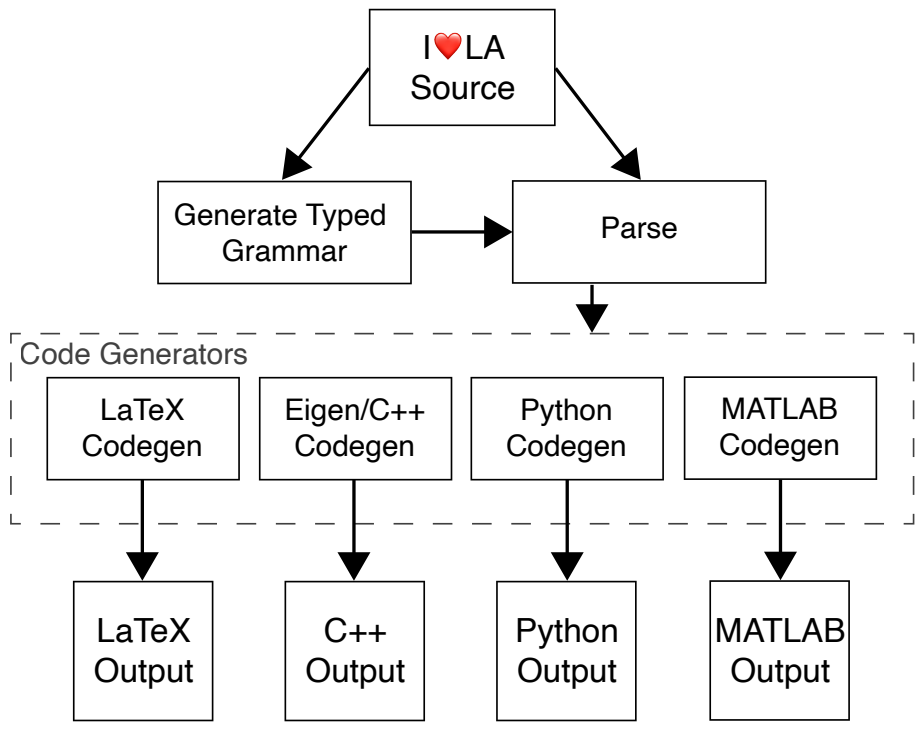

Fig. 3. Overview of the I♥LA compiler. A lightweight parse phase first examines the `given`, `where`, and `import` declarations, as well multi-letter variables that appear on the left-hand side of assignments, to determine the set of identifiers and their types. Then, the compiler parses the document into a simple intermediate form. Code generators handle conversion from the intermediate form to output source.

## 5 COMPILER

The I♥LA compiler implementation consists of approximately 10,000 lines of Python code and 900 lines of extended Backus-Naur form (EBNF) text specifying the grammar, built on top of the Tatsu [Añez 2019] parser generator library. While our EBNF is considerably longer than Python's approximately 500-line grammar[2], conventional math developed over time like a natural language, which is considerably more difficult to parse. The compiler uses the SymPy library to statically check compatibility of matrix and vector dimensions in the presence of block matrices and vectors, whose dimensions are sums of the dimensions of the individual blocks.

Fig. 3 shows an overview of how the compiler works. First, I♥LA source is parsed using an initial phase that parses `given` and `where` declarations, requests for built-in trigonometric or linear algebra functions, and multi-letter variables that appear on the left-hand side of variable definitions, to determine the set of identifiers used in the file as well as their types. Then, the compiler parses the I♥LA source into an intermediate representation (IR) that is suitable for translation to different output languages using output-specific code generators. The parser is not context-free, since the type and identifier information extracted during the initial parse is considered when parsing the I♥LA source into its IR.

We have written code generators for LaTeX (standalone and MathJax), C++ using Eigen, Python using NumPy/SciPy, and MATLAB; the latter three were chosen to highlight the ability to generate code for both statically-typed and dynamically-typed languages with 0- and 1-based indexing. The compiler is designed to be extended to support additional backend languages. Adding a new output language requires implementing a code generator class in Python, which walks over the IR and produces output source code. Currently, the C++/Eigen, Python/NumPy/SciPy, and MATLAB code generators consist of about 1300 lines each of Python code, while the LaTeX generator is only about 700 lines of code. We believe the relative succinctness of each backend demonstrates that only a small amount of effort is required to support additional backend languages.

Our compiler's second pass dynamically incorporates the identifier and type information extracted from the first pass. On average, over our test set of 168 I♥LA files, compilation of each I♥LA file to all four backends takes 0.2 seconds. The source code includes a lightweight GUI implemented using the wxPython cross-platform toolkit and a web GUI that runs the compiler in-browser using Pyodide. Both GUIs display the I♥LA source code, the compiled code in various output languages, and the rendered LaTeX output side-by-side. The source code editor performs ASCII-to-Unicode substitutions. The compiler and GUI run in a web browser or natively on macOS, Windows, and Linux.

[2]https://docs.python.org/3/reference/grammar.html

## 6 EVALUATION

To evaluate I♥LA and its expressiveness, we collected a set of formulas from the literature and a set of existing codebases. We describe our efforts to understand I♥LA's capabilities and limitations by implementing the collected formulas and comparing I♥LA *source code* to the original typeset formula in the PDF (Section 6.1), and by implementing formulas in I♥LA and using the generated output to replace functions in existing codebases (Section 6.2). We also conducted a statistical estimate of I♥LA's general applicability to equations appearing in computer graphics papers (Section 6.3). Section 7 reports on a user study we conducted.

### 6.1 Examples from the Wild

We demonstrate I♥LA's breadth with 26 representative equations selected from a variety of papers, books, and courses in computer graphics (Fig. 4). The supplemental materials contain a browsable gallery showing each original equation; its I♥LA implementation; the generated C++, Python, and LaTeX; and the typeset mathematics produced from the LaTeX (Fig. 5).

Of our examples, 12 are from *ACM SIGGRAPH* papers on image processing, geometric modeling, and animation [Blanz and Vetter 1999; Kim et al. 2019; Kondapaneni et al. 2019; Le and Lewis 2019; Liu et al. 2019a; McMillan and Bishop 1995; Rusinkiewicz 2019; Smith et al. 2019; Wang et al. 2019; Wronski et al. 2019], 7 are from a book on optimization [Boyd et al. 2004], 4 are from a geometry processing book [Botsch et al. 2010], and 3 are from a geometry processing course [Jacobson 2020].

These examples show that I♥LA's syntax is able to represent all of the operations with virtually identical symbols and only slightly degraded layout. A proportional font would decrease this gap, given the original LaTeX-produced math uses proportional fonts, though it would result in matrices with elements that are not aligned across different rows. Matrices in I♥LA lead to an undesirable vertical shift in the line of the expression, as can be seen for example in Fig. 4(o). I♥LA does support semicolons in place of line breaks, but this destroys the 2D visual layout of an individual matrix. We considered parsing matrices as 2D "ASCII art," but decided against this as the proper 2D layout of the rest of the line becomes ambiguous (e.g., should the rest of the formula continue to the right of the top or bottom row of a 2×2 matrix?) and it imposes a burden on the author to maintain a sensible 2D layout, since text must be edited on multiple rows consistently. For a similar reason, we did not explore 2D arrangements of fractions, either.

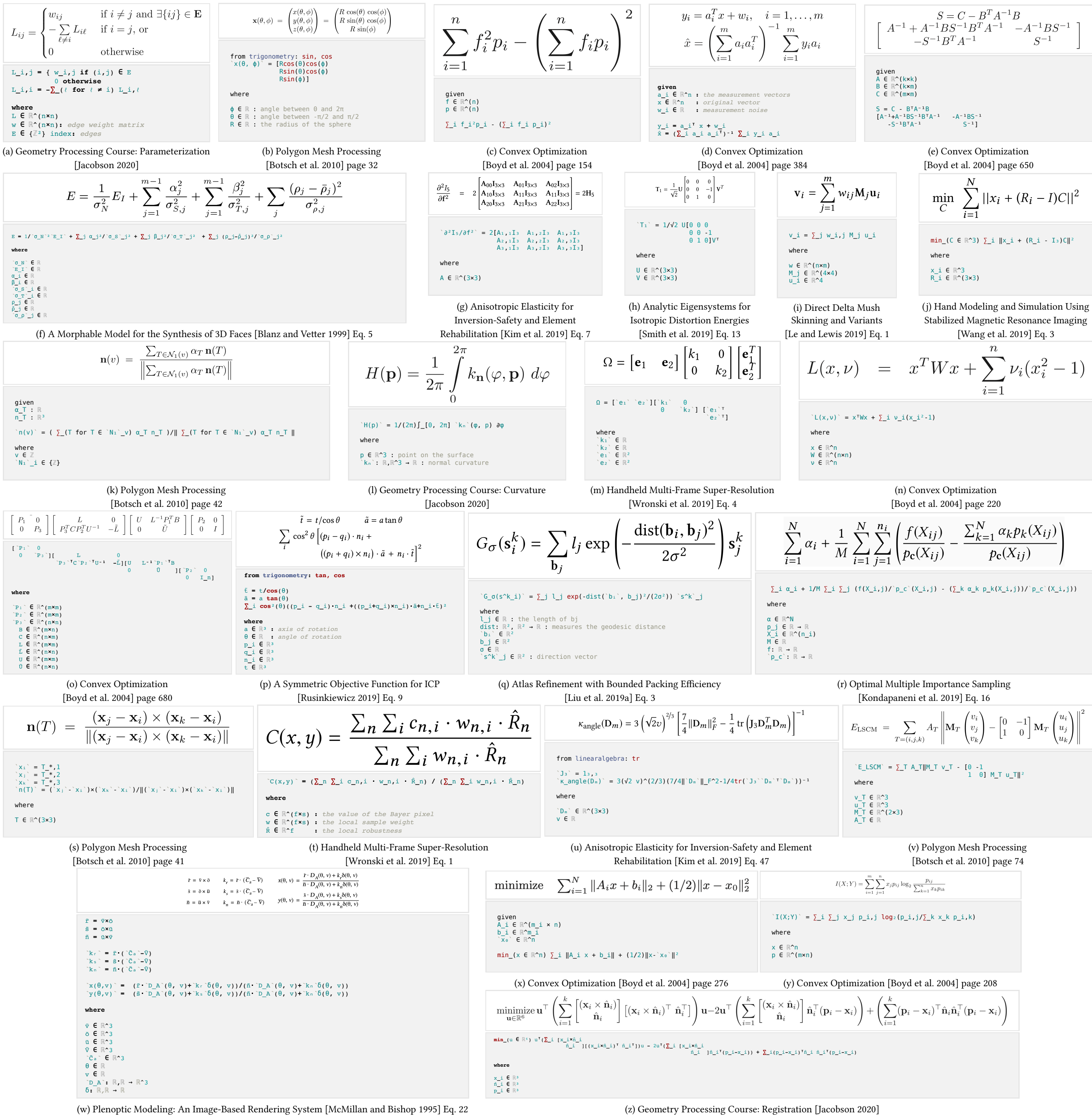

Fig. 4. Formulas from various papers, books, and courses in computer graphics and their translation into I❤LA.

These examples demonstrate certain limitations of I❤LA as a replacement for these equations in their original context. First, some equations define functions themselves, as in $E(x) = x^\top Px + q^\top x$. I❤LA code itself defines a function, so the "$E(x) =$" is redundant. To preserve the equation as written, we define a variable named "$E(x)$" as in `` `E(x)` = xᵀPx + qᵀx ``. We plan to explore local function definitions in the future, which would naturally provide support for this syntax as well as closures over precomputed values. More generally, some equations include more than one equals sign, such as $\nabla f = 2I_3 \otimes A = 2H$. Some equations mix definitions and derivations and thus don't follow I♥LA's new-symbol-on-the-left-hand side convention. I♥LA has no facility for such equations, though support could be added to I♥LA through new syntax that allows for commentary (e.g., derivations) on the same lines as definitions. I♥LA could also be expanded in the future with a computer algebra system to discover that, e.g., the unknown variable $H$ is defined by the expression $I_3 \otimes A$. This would add compile-time, but not run-time, complexity. I♥LA can only take linear algebra data types (matrices, vectors, sequences, sets, and functions) as input, rather than arbitrary or dynamic data structures. This requires us to choose a particular encoding for triangles or sliding windows (Figure 4 (k) and (t), respectively). In Figure 4 (u), I♥LA's LaTeX output does not typeset the $\kappa_{angle}$ subscript, because I♥LA escapes variable names in the LaTeX output. I♥LA can't assume that variable names are valid LaTeX. In this example, the variable name would have to be `` `κ_\text {angle}(D_m)` `` to obtain the desired typeset LaTeX, rather than `` `κ_angle(Dₘ)` ``.

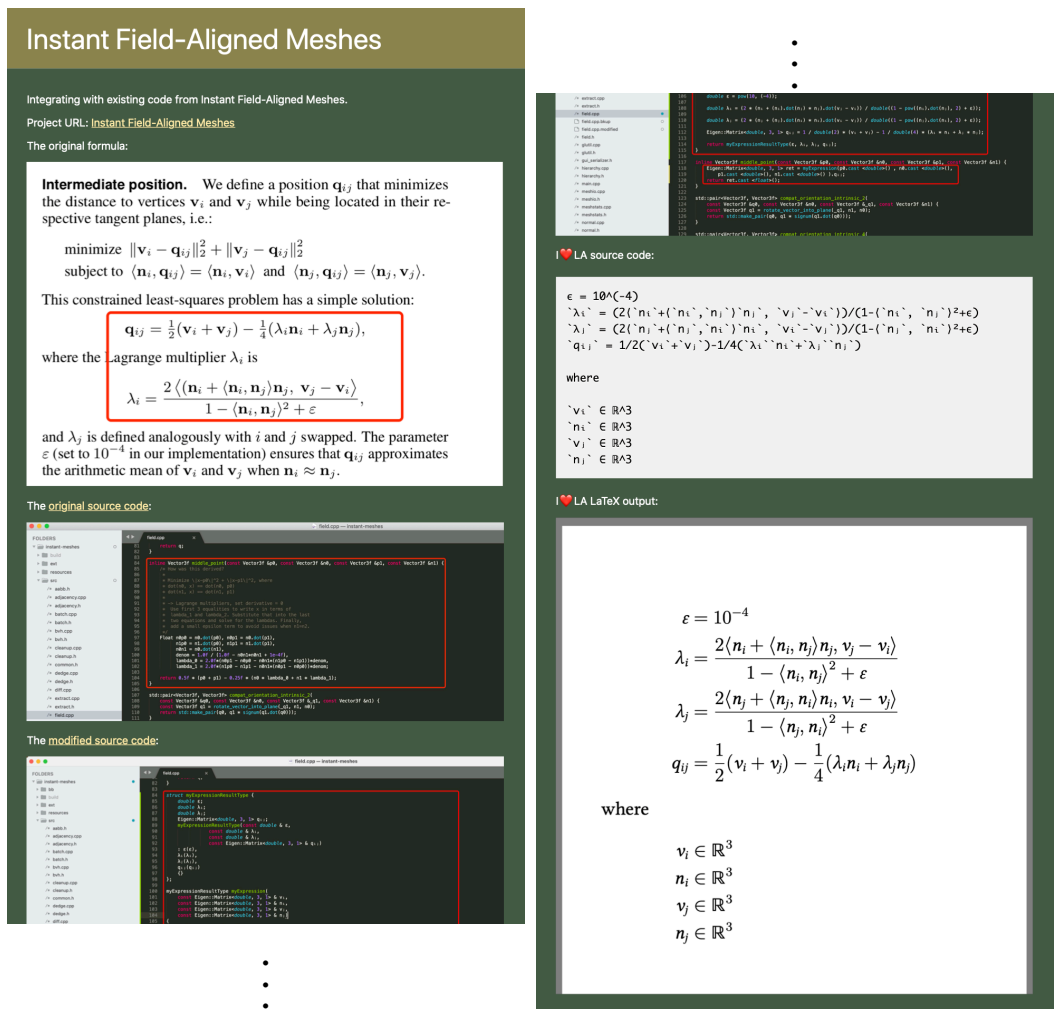

Fig. 5. The supplemental materials contain a page for each example in our evaluation.

## 6.2 Integrating with Existing Code

To simulate real-world I♥LA usage, we collected a set of existing codebases (paper implementations and libraries) and replaced functions within them with I♥LA implementations (Fig. 6). Note that not all code was associated with a formula in a paper written in mathematical notation; some formulas are just described in prose. Nevertheless, having an I♥LA expression for the code allows it to be instantiated in multiple programming environments and communicated clearly as LaTeX-formatted conventional math to other readers. We verified that replacing existing code with I♥LA-generated code produced identical output for each codebase. The case studies are summarized in Table 2. Please see the supplemental materials for the original and modified source code and the I♥LA source code (Fig. 5).

*Case Study 1: Robust Inside-Outside Segmentation using Generalized Winding Numbers.* This codebase is the original authors' implementation of a paper computing the inside-outside segmentations of shapes [Jacobson et al. 2013], and is implemented using C++ inside the libigl package [Jacobson et al. 2018]. We replaced the solid angle computation with I♥LA. The purpose of this function is difficult to discern based on the C++ code yet is easy to read in its I♥LA form (Fig. 7c). We show our output in Fig. 6.

*Case Study 2: Regularized Kelvinlets.* This codebase implements "sculpting brushes based on fundamental solutions of elasticity" [De Goes and James 2017] in C++, also as part of libigl. We can express the formula for the pinching operation ([De Goes and James 2017] Equation 17) as a single line of I♥LA similar to the way it is expressed in the paper (Fig. 7b). I♥LA's LaTeX output looks almost identical to the paper's formula, unlike the relatively difficult-to-read C++ code.

*Case Study 3: Instant Field-Aligned Meshes.* This codebase is the original authors' implementation of a remeshing algorithm [Jakob et al. 2015] in C++. We replace the "intermediate position" formula, which is the closed-form solution of a constrained least-squares problem. Again, 7 lines of hard-to-read C++ become 3–4 lines of I♥LA, which look very similar to the expression in the paper (Fig. 7a). We show their GUI running with I♥LA-generated code in Fig. 6.

*Case Study 4: Frame Fields.* We modify the original authors' implementation of a paper generating frame fields [Panozzo et al. 2014] using C++ in libigl. We replaced a canonical-to-frame transformation function with I♥LA. The function builds three matrices, of which the third is a block matrix involving the first two. This example highlights I♥LA's block matrix abilities. Though the formula involves statically-sized matrices, the developers use Eigen's flexible $n$-dimensional matrix type everywhere, whereas I♥LA's generated C++ code fills in Eigen's matrix dimension parameters precisely. This is typically tedious to do by hand. We show our code running Example 506 in the libigl tutorial in Fig. 6.

*Case Study 5: Collision-Aware and Online Compression of Rigid Body Simulations via Integrated Error Minimization.* This codebase is the authors' implementation of an animation-compression paper ([Jeruzalski et al. 2018]) in C++. We replaced a function computing the inverse of a block matrix (Equation 4 in the paper).

*Case Study 6: Properties of Laplace Operators for Tetrahedral Meshes.* This codebase is the original authors' C++ implementation of their paper on properties of volumetric Laplacians [Alexa et al. 2020]. We replaced two functions with I♥LA: the first function computes the volume of a tetrahedron, and the second calculates the circumcenter of a triangle. These functions are not complex. However, writing them in I♥LA produced a drop-in replacement for the existing function prototypes—and can also produce correct code in other languages.

Table 2. Summary of case studies. Among other benefits, I❤LA requires fewer lines of code (LoC) than the code it replaces.

| Case | Source | Language | LoC (original) | LoC (I❤LA) | Workarounds required |
|---|---|---|---|---|---|
| 1 | Jacobson et al. [2013] | C++ | 31 | 8 | No |
| 2 | De Goes and James [2017] | C++ | 6 | 1 | No |
| 3 | Jakob et al. [2015] | C++ | 7 | 4 | No |
| 4 | Panozzo et al. [2014] | C++ | 14 | 5 | No |
| 5 | Jeruzalski et al. [2018] | C++ | 5 | 2 | No |
| 6 | Alexa et al. [2020] | C++ | 12 | 6 | No |
| 7 | Sieger and Botsch [2020] | C++ | 26 | 9 | Yes |
| 8 | Alexa [2020] | C++ | 21 | 8 | Yes |
| 9 | Preiner et al. [2019] | Python | 15 | 9 | Yes |

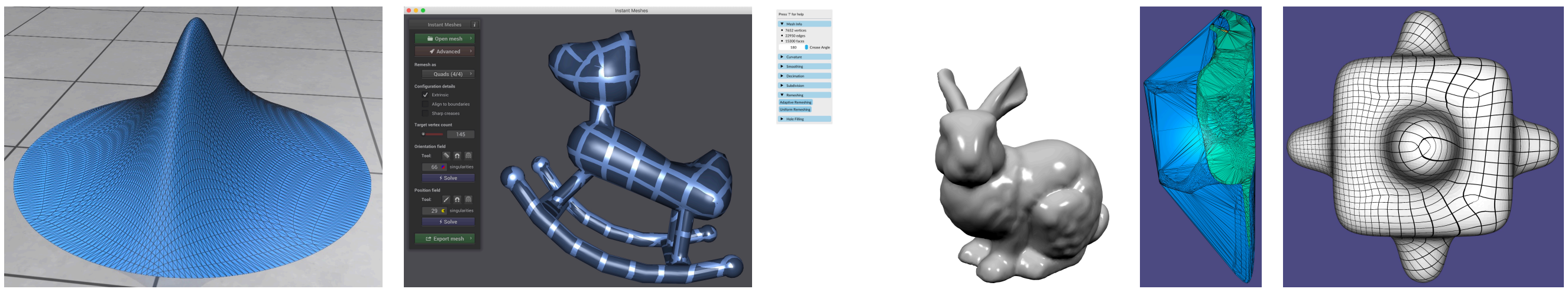

Fig. 6. We replaced code in existing code bases with code generated by our I❤LA compiler and verified that the output is identical. Here we show graphical examples. From left to right: Gaussian product subdivision surfaces [Preiner et al. 2019]; instant field-aligned meshes [Jakob et al. 2015]; Polygon Mesh Processing Library adaptive remeshing [Sieger and Botsch 2020]; winding number surface reconstruction [Jacobson et al. 2013]; frame fields [Panozzo et al. 2014].

*Case Study 7: Polygon Mesh Processing Library.* The Polygon Mesh Processing Library [Sieger and Botsch 2020] is written in C++. We replaced 26 lines of code implementing the Hessian and Jacobian matrices for an adaptive remeshing routine with I❤LA. The original C++ used some modular indexing arithmetic that I❤LA doesn't yet support, so we needed to create a copy of the matrix of indices with the arithmetic operations already applied. Fig. 6 shows the I❤LA-generated code running inside the library's GUI.

*Case Study 8: Conforming Weighted Delaunay Triangulations.* This codebase is the author's C++ implementation of their paper on Delaunay triangulation [Alexa 2020]. The author's code implements Equation 33 and part of Equation 34 in the paper, both of which we replaced with much shorter versions written in I❤LA. In order to interface with the existing code, we needed to add 3–4 lines of C++ code to pack the coordinates representing triangles into a matrix. In this example, a single I❤LA function returns 6 values consumed by the C++ code.

*Case Study 9: Gaussian-Product Subdivision Surfaces.* This codebase is the authors' Python implementation of their subdivision surface scheme [Preiner et al. 2019]. We replaced the code inside its key function with I❤LA. The function does not correspond to any formula in the paper, but is implemented in terms of linear algebra operations using NumPy. This core function makes use of an external library function with multiple return values, but I❤LA does not currently support calling external functions with multiple return values. To work around this, we copy-and-pasted the generated I❤LA code around the callback function. The callback function also changes the dimensions of an existing matrix, so we manually added a line of code to update I❤LA's dimension variable to match. The subdivided mesh can be seen in Fig. 6.

6.2.1 *Observations.* Overall, the I❤LA code is shorter and more closely matches the paper formula than the existing code it replaced. Importantly, it can also be compiled to multiple output languages.

Many codebases contain small formulas, e.g., `A = K.transpose() * (Minv * (-L * (Minv * K)))` [Jacobson et al. 2018], that could be expressed in conventional mathematical notation using I❤LA, e.g., $A = K^T M^{-1}(-L)M^{-1}K$. This suggests an embeddable or macro implementation of I❤LA would be beneficial. In a statically typed language, inline I❤LA could infer type declarations, eliminating the need to provide them in I❤LA.

I❤LA cannot replace the majority of code in a codebase. Formulas are ultimately a small part of an entire codebase, which includes file loading, documentation, GUIs, algorithms with complex control flow, and manipulation of complex data structures like meshes. This other code can often be re-used between projects, whereas the formulas in a new research paper must be accurately implemented anew in order to reproduce the work.

## 6.3 A Statistical Estimate of Applicability

To estimate the applicability of I❤LA to computer graphics *in general*, we randomly sampled 100 of the numbered equations from our corpus of all 1987 numbered equations appearing in the *ACM*

**Intermediate position.** We define a position $\mathbf{q}_{ij}$ that minimizes the distance to vertices $\mathbf{v}_i$ and $\mathbf{v}_j$ while being located in their respective tangent planes, i.e.:

$$\begin{aligned}&\text{minimize } \|\mathbf{v}_i - \mathbf{q}_{ij}\|_2^2 + \|\mathbf{v}_j - \mathbf{q}_{ij}\|_2^2\\ &\text{subject to } \langle \mathbf{n}_i, \mathbf{q}_{ij}\rangle = \langle \mathbf{n}_i, \mathbf{v}_i\rangle \text{ and } \langle \mathbf{n}_j, \mathbf{q}_{ij}\rangle = \langle \mathbf{n}_j, \mathbf{v}_j\rangle.\end{aligned}$$

This constrained least-squares problem has a simple solution:

$$\mathbf{q}_{ij} = \tfrac{1}{2}(\mathbf{v}_i + \mathbf{v}_j) - \tfrac{1}{4}(\lambda_i \mathbf{n}_i + \lambda_j \mathbf{n}_j),$$

where the Lagrange multiplier $\lambda_i$ is

$$\lambda_i = \frac{2\left\langle(\mathbf{n}_i + \langle\mathbf{n}_i, \mathbf{n}_j\rangle\mathbf{n}_j,\ \mathbf{v}_j - \mathbf{v}_i\right\rangle}{1 - \langle\mathbf{n}_i, \mathbf{n}_j\rangle^2 + \varepsilon},$$

and $\lambda_j$ is defined analogously with $i$ and $j$ swapped. The parameter $\varepsilon$ (set to $10^{-4}$ in our implementation) ensures that $\mathbf{q}_{ij}$ approximates the arithmetic mean of $\mathbf{v}_i$ and $\mathbf{v}_j$ when $\mathbf{n}_i \approx \mathbf{n}_j$.

```
ε = 10^(-4)
`λᵢ` = (2⟨`nᵢ`+⟨`nᵢ`,`nⱼ`⟩`nⱼ`, `vⱼ`-`vᵢ`⟩)/(1-⟨`nᵢ`, `nⱼ`⟩²+ε)
`λⱼ` = (2⟨`nⱼ`+⟨`nⱼ`,`nᵢ`⟩`nᵢ`, `vᵢ`-`vⱼ`⟩)/(1-⟨`nⱼ`, `nᵢ`⟩²+ε)
`qᵢⱼ` = 1/2(`vᵢ`+`vⱼ`)-1/4(`λᵢ``nᵢ`+`λⱼ``nⱼ`)

where

`vᵢ` ∈ ℝ^3
`nᵢ` ∈ ℝ^3
`vⱼ` ∈ ℝ^3
`nⱼ` ∈ ℝ^3
```

```
inline Vector3f middle_point(const Vector3f &p0, const Vector3f &n0, const Vector3f
↪ &p1, const Vector3f &n1) {
    Float n0p0 = n0.dot(p0), n0p1 = n0.dot(p1),
          n1p0 = n1.dot(p0), n1p1 = n1.dot(p1),
          n0n1 = n0.dot(n1),
          denom = 1.0f / (1.0f - n0n1*n0n1 + 1e-4f),
          lambda_0 = 2.0f*(n0p1 - n0p0 - n0n1*(n1p0 - n1p1))*denom,
          lambda_1 = 2.0f*(n1p0 - n1p1 - n0n1*(n0p1 - n0p0))*denom;

    return 0.5f * (p0 + p1) - 0.25f * (n0 * lambda_0 + n1 * lambda_1);
}
```

(a) Instant Field-Aligned Meshes [Jakob et al. 2015]

*Pinching:* The last type of locally affine regularized Kelvinlet is constructed based on a symmetric force matrix $\boldsymbol{F}$ with zero trace (in a total of 5 DoFs), and it generates displacements of the form

$$\boldsymbol{p}_\varepsilon(\boldsymbol{r}) = \frac{(2b-a)}{r_\varepsilon^3}\boldsymbol{F}\boldsymbol{r} - \frac{3}{2r_\varepsilon^5}\left[2\,b\,(\boldsymbol{r}^t\boldsymbol{F}\boldsymbol{r})\,\boldsymbol{I} + a\,\varepsilon^2\boldsymbol{F}\right]\boldsymbol{r}. \qquad (17)$$

```
`p_ε(r)` = (2b-a)/`r_ε`³ Fr - 3/(2`r_ε`⁵)(2b(rᵀFr)I_n+aε²F)r

where

F ∈ ℝ^(n×n)
r ∈ ℝ^n
`r_ε` ∈ ℝ
a ∈ ℝ
b ∈ ℝ
ε ∈ ℝ
```

```
// Regularized Kelvinlets: Formula (17)
kelvinlet = [&r, &F, &r_norm_sq, &kp](const Scalar& epsilon) {
  const auto r_epsilon = sqrt(r_norm_sq + kp.epsilon * kp.epsilon);
  const auto r_epsilon_3 = r_epsilon * r_epsilon * r_epsilon;
  auto t1 = ((2 * b - a) / r_epsilon_3) * F * r;
  auto t2_coeff = 3 / (2 * r_epsilon * r_epsilon * r_epsilon_3);
  auto t2 = t2_coeff * (2 * b * (r.transpose().dot(F * r)) * r +
                        a * epsilon * epsilon * epsilon * F * r);
  return t1 - t2;
};
```

(b) Regularized Kelvinlets [De Goes and James 2017]

And again, if we have a triangulated, piecewise-linear surface, there is an immediate and exact discretization of Equation (4):

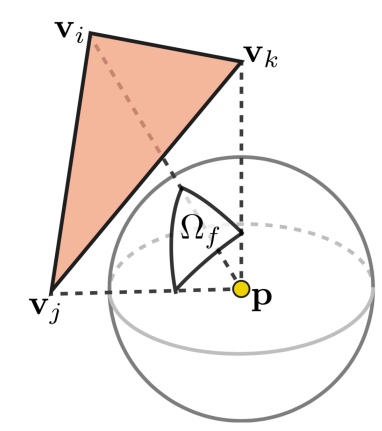

$$w(\mathbf{p}) = \sum_{f=1}^{m} \frac{1}{4\pi}\Omega_f(\mathbf{p}), \qquad (5)$$

where $\Omega_f$ is the solid angle of the **oriented** triangle $\{\mathbf{v}_i, \mathbf{v}_j, \mathbf{v}_k\}$ with respect to $\mathbf{p}$. Let $\mathbf{a} = \mathbf{v}_i - \mathbf{p}$, $\mathbf{b} = \mathbf{v}_j - \mathbf{p}$, $\mathbf{c} = \mathbf{v}_k - \mathbf{p}$ and $a = \|\mathbf{a}\|$, $b = \|\mathbf{b}\|$, $c = \|\mathbf{c}\|$; then following [van Oosterom and Strackee 1983]:

$$\tan\left(\frac{\Omega(\mathbf{p})}{2}\right) = \frac{\det([\mathbf{a}\,\mathbf{b}\,\mathbf{c}])}{abc + (\mathbf{a}\cdot\mathbf{b})c + (\mathbf{b}\cdot\mathbf{c})a + (\mathbf{c}\cdot\mathbf{a})b} \qquad (6)$$

```
from trigonometry: atan2

a = `vᵢ` - p
b = `vⱼ` - p
c = `vₖ` - p
a = ||a||
b = ||b||
c = ||c||

atan2(|[a b c]|, (abc+(a⋅b)c+(b⋅c)a+(c⋅a)b))/(2π)

where

`vᵢ` ∈ ℝ^3
`vⱼ` ∈ ℝ^3
`vₖ` ∈ ℝ^3
p ∈ ℝ^3
```

```
template <
  typename DerivedA,
  typename DerivedB,
  typename DerivedC,
  typename DerivedP>
IGL_INLINE typename DerivedA::Scalar
↪ igl::solid_angle(
  const Eigen::MatrixBase<DerivedA> & A,
  const Eigen::MatrixBase<DerivedB> & B,
  const Eigen::MatrixBase<DerivedC> & C,
  const Eigen::MatrixBase<DerivedP> & P)
{
  typedef typename DerivedA::Scalar SType;
  // Gather vectors to corners
  Eigen::Matrix<SType,3,3> v;
  // Don't use this since it will freak out for
  ↪ templates with != 3 size
  //v<< (A-P),(B-P),(C-P);
  for(int d = 0;d<3;d++)
  {
    v(0,d) = A(d)-P(d);
    v(1,d) = B(d)-P(d);
    v(2,d) = C(d)-P(d);
  }
  Eigen::Matrix<SType,1,3> vl = v.rowwise().norm();
  //printf("\n");
  // Compute determinant
  SType detf =
    v(0,0)*v(1,1)*v(2,2)+
    v(1,0)*v(2,1)*v(0,2)+
    v(2,0)*v(0,1)*v(1,2)-
    v(2,0)*v(1,1)*v(0,2)-
    v(1,0)*v(0,1)*v(2,2)-
    v(0,0)*v(2,1)*v(1,2);
  // Compute pairwise dotproducts
  Eigen::Matrix<SType,1,3> dp;
  dp(0) = v(1,0)*v(2,0);
  dp(0) += v(1,1)*v(2,1);
  dp(0) += v(1,2)*v(2,2);
  dp(1) = v(2,0)*v(0,0);
  dp(1) += v(2,1)*v(0,1);
  dp(1) += v(2,2)*v(0,2);
  dp(2) = v(0,0)*v(1,0);
  dp(2) += v(0,1)*v(1,1);
  dp(2) += v(0,2)*v(1,2);
  // Compute winding number
  // Only divide by TWO_PI instead of 4*pi because
  ↪ there was a 2 out front
  return atan2(detf,
    vl(0)*vl(1)*vl(2) +
    dp(0)*vl(0) +
    dp(1)*vl(1) +
    dp(2)*vl(2)) / (2.*igl::PI);
}
```

(c) Winding Number [Jacobson et al. 2013]

Fig. 7. Side-by-side comparisons of paper formula, the I♥LA implementation, and the original C++ implementation from three existing codebases (a,b,c).

*SIGGRAPH* 2019 proceedings. Among the 100 equations, we determined 15 to be derivations. Of the 85 remaining equations (formulas), we estimate that 53 (62%) are directly implementable using I❤LA while the remaining 32 are not (38%). Of the remaining 32, 11 rely on partial derivatives and gradients and unsupported integration, 10 express complex optimization problems, 7 use unsupported control flow, and 4 use unsupported operators. Some of the 11 equations using partial derivatives and gradients could be implemented in I❤LA by passing the partial derivatives and gradients as additional parameters. If so, this would allow I❤LA to implement $\frac{64}{85}$ = 75% of the formulas. Examples can be seen in Figure 8

## 7 USER STUDY

We conducted a user study to understand whether I❤LA can be learned and how it is perceived by experienced practitioners. We designed our experiment following the guidance of Ko et al. [2015]. We recruited 8 computer science PhD students from two universities based on their having at least six months experience implementing linear algebra formulas. 75% of participants reported at least one year of experience. Participants were familiar with a variety of numerical linear algebra programming environments. All participants reported spending over 1 hour a week programming (50% reported over 10 hours per week). In our experiment, participants implemented three progressively more difficult linear algebra formulas (Figure 9) in both I❤LA and their preferred programming environment (four used C++/Eigen and four used Python/NumPy/SciPy). Participants were given a test harness and skeleton code to use in their preferred programming environment. For I❤LA, participants used the browser-based editor and compiler. We counterbalanced experimental conditions so that half of the participants implemented all three formula using I❤LA first and half using their preferred environment first. Afterwards, participants filled out a post-study questionnaire. Participants received a $50 gift card as renumeration. Our protocol was approved by an institutional review board.

The study was conducted remotely in a one-on-one manner between an author and a participant. To learn I❤LA, participants were directed to the I❤LA language reference. Because I❤LA is a new language with few available resources, participants were also told to treat the author as an interactive language reference. Participants were told to spent at most 10 minutes for each implementation, though we allowed participants to use more time if desired. All participants chose to work until successful completion, which took no more than 21 minutes for each task. The experiment lasted between one and two hours for each participant.

Raw data and survey responses can be found in the supplemental materials. We asked three questions to assess participants' perceptions about I❤LA on a 5-point Likert scale ranging from 1 (strongly disagree) to 5 (strongly agree). This data is shown in Figure 10. We computed $p$-values using a one-sample non-parametric permutation test [Raschka 2018] against the 'neutral' theoretical median response and then corrected them using a post-hoc Holm-Bonferroni familywise error correction [Seabold and Perktold 2010]. All three questions achieved statistical significance ($p < 0.01$). Our experiment found that participants found I❤LA easy to learn and that I❤LA looks similar to conventional math. Most intriguingly, participants preferred I❤LA to the other language they spent months and years using.

| | | |
|---|---|---|
| Unsupported Operators | $\Sigma_i^v = \operatorname{cov}(v_i \cup \mathcal{N}(i)) + \sigma_0^2 I$ | [Preiner et al. 2019] |
| | $0 \le \begin{pmatrix} \mathbf{A} & \mathbf{J}^T & \mathbf{J}_t^T & 0 \\ \mathbf{J} & 0 & 0 & 0 \\ \mathbf{J}_t & 0 & 0 & -\mathbf{e}^T \\ 0 & \mu & -\mathbf{e} & 0 \end{pmatrix} \begin{pmatrix} \mathbf{v} \\ \lambda \\ \gamma \\ \beta \end{pmatrix} - \begin{pmatrix} \mathbf{b} \\ \mathbf{c}_n \\ \mathbf{c}_t \\ 0 \end{pmatrix} \perp \begin{pmatrix} 1 \\ \lambda \\ \gamma \\ \beta \end{pmatrix} \ge 0$ | [Verschoor and Jalba 2019] |
| Derivations | $\frac{C}{C^0} = \frac{\epsilon_r \epsilon_0 \frac{lw}{d}}{\epsilon_r \epsilon_0 \frac{l^0 w^0}{d^0}} = \frac{l}{l^0} \frac{w}{w^0} \frac{d^0}{d} = \frac{l}{l^0}$ | [Glauser et al. 2019] |
| | $\sum_{j=1}^{n} P_{i,j} \le f_0(x_i), \forall i$ | [Bonneel and Coeurjolly 2019] |
| Unsupported Optimization | $\underset{\widetilde{\mathsf{L}} \doteq \mathsf{S}_{\widetilde{\mathsf{L}}}}{\text{minimize}} \underbrace{\frac{1}{2} \Vert \mathsf{P}\mathsf{M}^{-1}\mathsf{L}\Phi - \widetilde{\mathsf{M}}^{-1}\widetilde{\mathsf{L}}\mathsf{P}\Phi \Vert_{\widetilde{\mathsf{M}}}^2}_{E_k(\widetilde{\mathsf{L}})}$ subject to $\widetilde{\mathsf{L}}$ is positive semi-definite and $\widetilde{\mathsf{L}}\mathsf{P}\Phi_0 = 0$ | [Liu et al. 2019b] |
| | $\mu(\mathbf{X}) = \max_{1 \le i \ne j \le n} \frac{\lvert \mathbf{X}_{.,i}^T \mathbf{X}_{.,j} \rvert}{\Vert \mathbf{X}_{.,i} \Vert_2 \Vert \mathbf{X}_{.,j} \Vert_2}$ | [Miandji et al. 2019] |
| | $\min_{c_0, c_2: I \to \mathbb{C}} E_{\text{alignment}} + \lambda E_{\text{smoothness}} + \mu E_{\text{regularization}} = \min_{c_0, c_2: I \to \mathbb{C}} \int_I \lvert f(e^{i\theta_\tau}; c_0(\vec{x}), c_2(\vec{x})) \rvert^2 d\vec{x} + \lambda \sum_{i=0,2} \int_I \Vert \nabla c_i(\vec{x}) \Vert^2 d\vec{x} + \mu \int_I \lvert f(e^{i\theta_g}; c_0(\vec{x}), c_2(\vec{x})) \rvert^2 d\vec{x}$ | [Bessmeltsev and Solomon 2019] |
| | $\min_{\mathbf{X}, \overline{\mathbf{X}}, \mathbf{d}} \; w_d M_d + w_p M_p + w_l M_l + w_s M_s + w_e M_e$ s.t. $\mathbf{f}_m(\mathbf{X}, \overline{\mathbf{X}}) + \mathbf{f}_p(\overline{\mathbf{X}}, \mathbf{d}) + \mathbf{f}_e(\overline{\mathbf{X}}) = 0$, $\overline{\mathbf{x}}_i = \mathbf{s}_i \quad (\forall \mathbf{s}_i \in \Psi)$ | [Zhang et al. 2019] |
| Multiple Conditions | $w_{\text{avr}}(q) = \frac{1}{\pi} \begin{cases} \frac{1}{40}(15q^3 - 36q^2 + 40) & 0 \le q < 1, \\ \frac{-3}{4q^3}\left(\frac{q^6}{6} - \frac{6q^5}{5} + 3q^4 - \frac{8q^3}{3} + \frac{1}{15}\right) & 1 \le q < 2, \\ \frac{3}{4q^3} & q \ge 2. \end{cases}$ | [Huang et al. 2019] |
| Other | $(\mathsf{L}\phi)_i = \frac{1}{2} \sum_{ij \in E} \underbrace{(\cot \theta_k^{ij} + \cot \theta_l^{ji})}_{=: w_{ij}} (\phi_j - \phi_i)$ | [Sharp et al. 2019b] |
| | $\Psi(x,y,z,t) = \iiint \Phi(k_x, k_y, k_z) e^{2\pi i (k_x x + k_y y + k_z z - f t)} \, \mathrm{d}k_x \mathrm{d}k_y \mathrm{d}k_z$ | [Lindell et al. 2019] |
| Derivatives | $J_{\text{tv}}(\mathbf{V}_\alpha) = \frac{1}{N} \sum_{\mathbf{x}} \lambda_{\text{tv}} \left\Vert \frac{\partial}{\partial \mathbf{x}} \log \mathbf{V}_\alpha(\mathbf{x}) \right\Vert$ | [Lombardi et al. 2019] |
| | $\lambda_{1,2} = 2 \frac{\partial \Psi_5}{\partial I_5}$ | [Kim et al. 2019] |

Fig. 8. Example equations that can't be directly implemented in I❤LA. The derivative expressions could be implemented by passing the derivatives as parameters.

We also compared the time it took participants to implement the formulas using I❤LA and participants' other language (Table 3). When implementing the simple formulas—also when participants' first encountered and learned to use I❤LA—they completed the task

**Simple**

Given an $n \times n$ matrix $A$, an $n$-vector $b$, and a constant $c$, compute the quadratic form for an $n$-vector $x$:

$$x^T A x + b^T x + c$$

**Medium**

Multiply a 3D vertex position $v$ by a weighted average of $4 \times 4$ transformation matrices $T_i$. The corresponding weights are $w_i$. Assume the vertex position $v$ is already in homogeneous coordinates, which is to say $v$ is a 4-vector.

$$u = \sum_i w_i T_i v$$

**Complex**

Create an edge-weighted adjacency matrix. Given a set of edges $E$ for a graph of $n$ vertices $v_i$, create the matrix:

$$A_{ij} = \begin{cases} \frac{1}{\|v_i - v_j\|} & \text{if } i,j \in E \\ 0 & \text{otherwise} \end{cases}$$

Fig. 9. The simple, medium, and complex tasks in our user study. The complex task required participants to create a sparse matrix.

Table 3. The average time in minutes participants needed to complete each programming task using I♥LA and their other programming environment (C++/Eigen or Python/NumPy/SciPy). Averages are rounded to the nearest minute. Statistical significance ($p$-values) have been adjusted with a familywise error correction; significant values are shown in bold.

| | simple | medium | complex |
|---|---|---|---|
| I♥LA (minutes) | 10 | 9 | 12 |
| Other (minutes) | 4 | 6 | 12 |
| Significance ($p$) | **0.005** | 0.065 | 0.862 |

faster using the other language they are experienced in. This difference decreased for the later tasks. Using a paired t-test and a post-hoc Holm-Bonferroni familywise error correction, only the averaged timings for the simple formula was statistically significant. We find these times extremely encouraging. Over the course of our short experiment, participants were able to learn I♥LA *and* complete tasks as efficiently as in their experienced language.

One participant commented, "I liked that it felt like a combination of markdown and latex. It's so nice to type so few lines compared to my c++ implementation. There is so much time and space used declaring the correct variables, initializing matrices to zero, figuring out how to index things, unnecessary loops. I like how short the I♥LA code was. I feel it allows to think about the math directly and skip implementation part and finding/fixing bugs too. Usually those bugs in c++ are annoying and a distraction from the theory." Another participant stated, "I heart LA is easier to implement comparing to Python even for I heart LA beginners. I feel like I don't even need to 'program' the I heart LA to get the formula function that I want for other programming languages." Regarding our choice of using Unicode symbols, one participant remarked, "I was initially worried that I would have to learn how to enter those characters from my keyboard, but the ASCII commands that convert to unicode in the web editor were very useful, so I think this is basically a non-issue with proper editor support."

We conclude from our user study that users can accomplish a range of tasks in I♥LA within 15 minutes and that users perceive that I♥LA looks similar to conventional math.

## 8 CONCLUSION

I♥LA has the potential to greatly benefit the scientific ecosystem, from researchers and practitioners to students and teachers. I♥LA makes it easier to try new ideas, since it automates the tedious and error-prone step of translating a mathematical idea into compilable code. I♥LA makes it easier to publish ideas correctly, since the published formula can be—or will have been—compiled and tested. I♥LA makes it easier to reproduce ideas, since mathematical notation changes relatively infrequently while programming environments change frequently. I♥LA may reduce the *translation loss* that occurs when communicating via publications: A graduate student describes their implementation to a co-author who writes it in LaTeX, introducing an error; then, someone else reads the paper and implements it incorrectly.

As Bonneel et al. [2020] found, often code released with papers can be difficult to compile and run in an environment different from the authors. Static environments are elusive. Operating systems, compilers, and support libraries are constantly being updated and sometimes break compatibility. Even when the code compiles and executes, someone may wish to make use of only a small portion of the authors' code. This can be difficult to extract due to poor or mismatched encapsulation. Submitting papers with verified I♥LA guarantees a baseline level of reproducibility.

The existence of I♥LA raises interesting pedagogical questions. Consider, for example, a lecture-based course where materials are delivered via conventional math on the chalkboard, and then students demonstrate their learning by *implementing* this math in C++ code. How much of the students' learning activity is an exercise in *translation*? If the target language is changed to I♥LA and translating conventional math to code becomes trivial, then how should such lecture-and-assignment-based courses adapt?

Unicode symbols are very beneficial to the readability of I♥LA. Conventional mathematical symbols are a core part of conventional math. Via Unicode, I♥LA strongly resembles conventional math (even without compiling and rendering LaTeX). Unicode is no harder to type than LaTeX symbol names or the long operator names in conventional programming languages. We started I♥LA with the conceit that variable names would be single letters. Although we relaxed this limitation, I♥LA—just like conventional math—is most readable when using single-letter identifiers.

*Limitations and Future Work.* I♥LA is deliberately limited in scope. We do not wish for I♥LA to become a general purpose programming language, but rather to complement existing programming ecosystems and improve access to correctness. We do not wish for I♥LA to produce code that runs at state-of-the-art efficiency, but it could facilitate adoption of new high-performance linear algebra packages with the addition of new code generators. We plan to add more target languages and facilitate others' to add their own code generators. We also plan to expand I♥LA's syntax with useful, unambiguous conventions as needs arise.

Mathematical notation is vast and redundant [Tao 2020]. Our scope is limited to linear algebra as expressed in computer graphics. While our examples (Section 6.1) also draw from a book on optimization [Boyd et al. 2004] and we know from experience that the linear algebra notation used in computer graphics is compatible with the notation found elsewhere, we did not conduct systematic analyses on equations from other domains. We expect that I❤LA will be applicable to adjacent fields like computer vision, machine learning, robotics, and optimization. Case studies will likely reveal additional notation to support, such as probabilistic expressions. It is worth considering other flavors of mathematical notation in the future, as an extension of I❤LA or as a similarly-inspired independent language. For example, Einstein notation is popular in some domains, but virtually no SIGGRAPH papers use it (none in 2019).

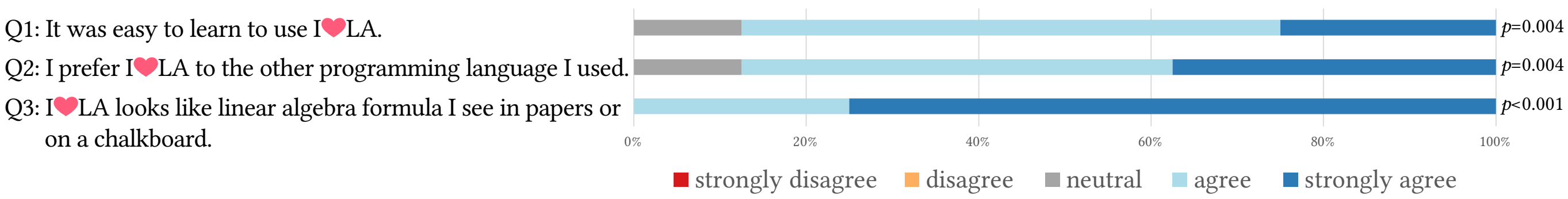

Fig. 10. Qualitative data from our user study.

We don't yet support creating local functions, calling functions with multiple return values like singular value or Eigen-decomposition, arithmetic in subscripts, such as `a_(2i)`, complex numbers and quaternions, or tensors. We started with the most common notation based on our quantitative analysis of SIGGRAPH equations. Subscript arithmetic is present in 5.7% (113/1987) of the equations; complex numbers and quaternions in 2.1% (41/1987); multiple returns values in 0.5% (10/1987); and Eigen and singular value decompositions in 0.4% (8/1987). Local functions are rarely relevant at the level of individual equations.

We plan to explore control flow in future extensions to I❤LA. Pseudocode in papers sometimes contains mistakes; pseudocode is an *ad hoc* invented language that cannot be compiled and tested. Even for papers that include original codebases, pseudocode provides a clear recipe for implementing the key ideas independent of any one programming language.

I❤LA's intermediate representation (IR) is math. It could be useful for symbolic manipulations like derivatives and distributed computation. Separating the IR into a standalone artifact could allow for re-usability benefits as in the successful LLVM project [Lattner and Adve 2004]. The IR could be compiled to various high-performance outputs. Other languages could target the IR and gain access to the compilers. It would be extremely beneficial to solve the inverse problem of converting existing codebases to I❤LA by first lifting into the IR as in [Ahmad et al. 2019]. The I❤LA code could be used as a more readable version of the function and as a means to convert the code into a new language. Along these lines, we plan to write code generators for additional programming languages and mathematical modeling environments. It would also be interesting to consider the problem of converting hand-written math to I❤LA.

More broadly, we are interested in extending I❤LA into a literate environment so that prose and formula can be interwoven. This could take the form of a new type of cell for Jupyter notebooks that both displays a typeset formula (via LaTeX output) and can be executed (like cells containing code). This could also take the form of an environment for writing executable academic papers.

Finally, we are interested in additional user studies. The study we performed shows that researchers can become proficient in I❤LA quite quickly. A longitudinal study could investigate whether researchers who use I❤LA are able to test more ideas.

## ACKNOWLEDGMENTS

We are grateful to the anonymous reviewers for their suggestions, Towaki Takikawa for helpful feedback, and Thomas LaToza for a discussion on evaluating programming languages. Alec Jacobson was supported in part by the Canada Research Chairs Program Yotam Gingold was supported in part by the United States National Science Foundation (IIS-1453018) and a gift from Adobe Systems Inc.